\documentclass[12pt]{article}
\begin{document}
\title{Modelling the linear viscoelastic behavior 
of silicate glasses near the glass transition point}
\author{Aleksey D. Drozdov and Jesper deC. Christiansen\\
Department of Production\\ 
Aalborg University\\
Fibigerstraede 16, DK--9220 Aalborg, Denmark}
\date{}
\maketitle

\begin{abstract}
A model is derived for the viscoelastic response of glasses
at isothermal uniaxial deformation with small strains.
A glass is treated as an ensemble of relaxing units 
with various activation energies for rearrangement.
With reference to the energy-landscape concept,
the rearrangement process is thought of as a series of hops of relaxing units
(trapped in their potential wells on the energy landscape) to higher
energy levels.
Stress--strain relations are developed by using the laws of
thermodynamics.
Adjustable parameters are found by fitting experimental data 
in torsional dynamic tests on a multicomponent silicate glass 
at several temperatures near the glass transition point.
\end{abstract}
\vspace*{5 mm}

\noindent
{\bf Key-words:} Silicate glasses, Dynamic tests, Viscoelasticity,
Glass transition

\section{Introduction}

This study is concerned with the viscoelastic behavior of 
silicate glasses at uniaxial deformation with small strains.
This subject has been a focus of attention in the past decade
because of numerous applications of vitreous silica 
in industry (that range from waveguides and optical fibers 
to semiconductor wafers and tire additives \cite{Fan91}
and the key role of multicomponent silicate glasses 
in geological processes \cite{Mys88}.

Observations in static and dynamic mechanical tests
on silicate glasses are conventionally fitted by using
such phenomenological approaches as 
(i) the generalized Maxwell model \cite{Bri96},
(ii) the KWW (Kolhrausch--Williams--Watt) formula \cite{DGB97},
and (iii) linear stress--strain relations with fractional derivatives \cite{PPD00}.
The objective of this paper is to develop constitutive equations
for the linear viscoelastic response of an amorphous glass based on
a micro-mechanical concept and to validate these relations by matching
experimental data in torsional oscillatory tests on a multicomponent silicate glass.

\section{A micro-mechanical model}

A silicate glass is thought of as an ensemble of cooperatively
rearranging regions (CRRs), where mechanical stress relaxes
due to rotational rearrangement of small clusters of atoms 
(consisting of a few SiO$_{4}$ tetrahedra)
that change their configuration as they are agitated by thermal fluctuations.

To describe the viscoelastic response of a glass in dynamic tests 
with a restricted window of frequencies, we distinguish two types of relaxing units:
(i) active CRRs that can rearrange during the experimental time-scale,
and (ii) passive CRRs whose characteristic time for rearrangement
substantially exceeds the duration of a test.
An active CRR is entirely characterized by the energy, $v$, of
thermal fluctuations necessary for rearrangement.
With reference to the energy-landscape concept \cite{Gol69},
an active CRR is modelled as a material point trapped in a potential well 
on the energy hypersurface in the phase space.
The well is characterized by its depth, $v$, with respect
to some reference (liquid-like) energy level.
The point spends most time being located at the bottom level 
of its potential well.
At random instants, the point hops to higher energy levels 
as it is thermally agitated.
If the point does not reach the liquid-like energy level in
a hop, it returns to the bottom level of its well being unaltered.
when the point reaches the reference level in a hop, 
the CRR changes its configuration and accepts the current state 
of a deformed medium as its reference state.

The kinetics of rearrangement is uniquely determined by
the temperature-dependent attempt rate $\Gamma_{0}$ 
(the number of hops in a well per unit time) 
and the distribution function, $q(u)$,
for hops with various heights ($q(u)du$ is the probability of a hop
whose height belongs to the interval $[u, u+du ]$).
We adopt the exponential distribution of intensities of hops \cite{Bou92},
\[ 
q(u)=\frac{1}{V}\exp \Bigl (-\frac{u}{V} \Bigr )\quad
(u\geq 0),
\]
where $V$ is the average height to be reached in a hop.
For a CRR trapped in a well with depth $v$, the probability 
of rearrangement in a hop reads
\[
Q(v)=\int_{v}^{\infty} q(u) du =\exp (-\frac{v}{V} ).
\]
The rate of rearrangement, 
\[
\Gamma(v)= \Gamma_{0}Q(v), 
\]
equals the number of hops (per unit time) in which 
a point reaches the liquid-like level,
\begin{equation}
\Gamma(v)= \Gamma_{0}\exp \Bigl (-\frac{v}{V}\Bigr ).
\end{equation}
Denote by $N_{\rm a}$ and $N_{\rm p}$ the numbers of active 
and passive CRRs per unit mass, and by $p(v)$ the distribution 
of potential wells with various energies $v$.
The rearrangement process is described by the function 
$n(t,\tau,v)$ that equals the number (per unit mass) 
of active CRRs trapped in wells with energy $v$ 
which have last been rearranged before instant $\tau\in [0,t]$.
The quantity $n(t,t,v)$ is the current number of active 
CRRs (per unit mass) with energy $v$,
\begin{equation}
n(t,t,v)=N_{\rm a}p(v).
\end{equation}
The quantity
\[
\frac{\partial n}{\partial \tau}(t,\tau,v)d\tau
\]
equals the number (per unit mass) of active CRRs with energy $v$
that have last been rearranged during the interval $[\tau,\tau+d\tau ]$ 
and, afterwards, have not beed rearranged within the interval $[\tau,t]$.
The number of active CRRs (per unit mass) with energy $v$ that were 
rearranged (for the first time) within the interval $[t,t+dt ]$ is given by
\[
-\frac{\partial n}{\partial t}(t,0,v)dt,
\]
and the number (per unit mass) of active CRRs with energy $v$ 
that have been rearranged within the interval $[\tau,\tau+d\tau ]$,
and, afterwards, were rearranged during the interval $[t,t+dt ]$
reads
\[
-\frac{\partial^{2} n}{\partial t\partial \tau }(t,\tau,v)dt\: d\tau.
\]
The rate of rearrangement, $\Gamma$, is defined as the ratio of
the number of active CRRs rearranged per unit time 
to the total number of active CRRs,
\begin{eqnarray*}
\frac{\partial n}{\partial t}(t,0,v) &=& -\Gamma(v) n(t,0,v),
\nonumber\\
\frac{\partial ^{2} n}{\partial t\partial \tau}(t,\tau,v) &=&
-\Gamma(v)\frac{\partial n}{\partial\tau }(t,\tau,v).
\end{eqnarray*}
The solutions of these equations with initial condition (2) read
\begin{eqnarray}
n(t,0,v) &=& N_{\rm a}p(v)\exp \Bigl [-\Gamma(v)t\Bigr ],
\nonumber\\
\frac{\partial n}{\partial \tau}(t,\tau,v) &=& 
N_{\rm a}p(v)\Gamma(v) \exp \Bigl [ -\Gamma(v)(t-\tau) \Bigr ].
\end{eqnarray}

\section{Constitutive equations}

At uniaxial deformation with small strains, 
a CRR is modelled as a linear elastic medium 
with the strain energy
\[
w=\frac{1}{2} \mu e^{2},
\]
where $\mu$ is an average rigidity per CRR
and $e$ is the strain from the reference
state of a relaxing unit to its deformed state.
We assume that CRRs are connected by links that
transmit the macro-strain in an ensemble, $\epsilon$, 
to individual regions.
This means that for a passive CRR (that is not
rearranged within the experimental time-scale), as well as
for an active CRR that has not been rearranged 
before time $t$, the strain $e$ coincides with 
the macro-strain $\epsilon(t)$.
For a CRR that has last been rearranged at time $\tau\in [0,t]$, 
the reference state coincides with the state of 
the ensemble at the instant of rearrangement,
which implies that the strain, $e$, equals 
the difference between the current macro-strain, $\epsilon(t)$, 
and the macro-strain, $\epsilon(\tau)$, at the instant 
of rearrangement, $\tau$.

To calculate the strain energy density per unit mass, $W(t)$, 
we multiply the mechanical energy per CRR by the number of 
relaxing units and sum the mechanical energies of passive CRRs 
and active CRRs rearranged at various instants $\tau\in [0,t]$.
Neglecting the energy of interaction between relaxing units, 
we find that
\begin{eqnarray}
W(t) &=& \frac{1}{2}\mu \biggl \{ \biggl [ N_{\rm p} 
+\int_{0}^{\infty} n(t,0,v)dv \biggr ] \epsilon^{2}(t) 
\nonumber\\
&&+ \int_{0}^{\infty} dv 
\int_{0}^{t} \frac{\partial n}{\partial \tau}(t,\tau,v) 
\Bigl [ \epsilon(t)-\epsilon(\tau)\Bigr ]^{2} d\tau \biggr \} .
\end{eqnarray}
Differentiation of Eq. (4) with respect to time results in
\begin{eqnarray}
&& \frac{dW}{dt}(t) = A(t)\frac{d\epsilon}{dt}(t)-Y(t),
\\
&& A(t)= \mu N \biggl [ \epsilon(t)-\kappa
\int_{0}^{t} \epsilon(\tau) d\tau \int_{0}^{\infty} \Gamma(v)
\exp \Bigl (-\Gamma(v)(t-\tau)\Bigr ) p(v) dv \biggr ],
\\
&& Y(t)= \frac{1}{2}\mu \int_{0}^{\infty} \Gamma(v) dv
\biggl [ n(t,0,v) \epsilon^{2}(t)  
+ \int_{0}^{t} \frac{\partial n}{\partial \tau}(t,\tau,v) 
\Bigl ( \epsilon(t)-\epsilon(\tau)\Bigr )^{2}  d\tau \biggr ],
\end{eqnarray}
where $N=N_{\rm a}+N_{\rm p}$ and $\kappa=N_{\rm a}/N$.
For isothermal uniaxial deformation, the Clausius--Duhem 
inequality reads
\[
\Theta=-\frac{dW}{dt}+\frac{\sigma}{\rho}\frac{d\epsilon}{dt} \geq 0,
\]
where $\rho$ is mass density, 
$\sigma$ is stress,
and $\Theta$ is internal dissipation per unit mass. 
Substitution of Eq. (5) into this equation yields
\[
\Theta=\frac{1}{\rho} (\sigma-\rho A)\frac{d\epsilon}{dt}+Y\geq 0.
\]
It follows from Eq. (7) that $Y(t)$ is non-negative.
This implies that the dissipation inequality is satisfied, 
provided that the first term on the right-hand side vanishes.
Equating the expression in brackets to zero and using Eq. (6), we 
arrive at the constitutive equation
\begin{equation}
\sigma(t) = G \biggl [ \epsilon(t)-\kappa \int_{0}^{t} \epsilon(\tau) d\tau
\int_{0}^{\infty} \Gamma(v)
\exp \Bigl (-\Gamma(v)(t-\tau)\Bigr ) p(v) dv \biggr ],
\end{equation}
where $G=\rho \mu N$ is an elastic modulus.

\section{Dynamic mechanical tests}

For an oscillatory test with an amplitude $\epsilon_{0}$ and a frequency $\omega$,
the strain reads 
\[
\epsilon(t)=\epsilon_{0}\exp (i\omega t), 
\]
where $i=\sqrt{-1}$.
Substituting this expression into Eq. (8) and using Eq. (1),
we find the storage, $G^{\prime}(\omega)$, 
and loss, $G^{\prime\prime}(\omega)$, moduli 
\begin{eqnarray}
G^{\prime}(\omega) &=& G \Bigl [ 1 -\kappa_{0}
\int_{0}^{\infty} \frac{\Gamma_{0}^{2}\exp(-2z)}{\Gamma_{0}^{2}\exp(-2z)+\omega^{2}}
P(z) dz \Bigr ],
\nonumber\\
G^{\prime\prime}(\omega) &=& G\kappa_{0}\omega \int_{0}^{\infty} 
\frac{\Gamma_{0}\exp(-z)}{\Gamma_{0}^{2}\exp(-2z)+\omega^{2}} P(z) dz,
\end{eqnarray}
where 
\[
z=\frac{v}{V},
\qquad
\kappa_{0}=\kappa V,
\qquad
P(z)=p(Vz).
\]
Restrictions on the upper boundary of a frequency window in conventional
mechanical tests (the maximal frequency does not exceed 1 kHz)
imply that the shape of the distribution function $P(z)$ at small $z$ 
cannot be uniquely determined by fitting observations.
To avoid ambiguities, we set 
\begin{equation}
P(z)=K_{1}\delta(z)+K_{2}P_{0}(z),
\end{equation}
where $\delta(z)$ is the Dirac delta-function, 
$K_{m}$ ($m=1,2$) are non-negative constants such that $K_{1}+K_{2}=1$,
and $P_{0}(z)$ is the distribution function corresponding 
to the range of frequencies ``visible" in the frequency window.
With reference to \cite{WSN01}, the following expression is adopted 
for the function $P_{0}(z)$:
\begin{equation}
P_{0}(z)=\frac{1}{6Z} \Bigl (\frac{z}{Z}\Bigr )^{3} \exp \Bigl (-\frac{z}{Z} \Bigr ),
\end{equation}
where 
\[ Z=\frac{\langle z\rangle}{4}
\]
and $\langle z\rangle$ is the average activation energy per CRR.
Substitution of Eq. (10) into Eq. (9) results in
\begin{eqnarray}
G^{\prime}(\omega) &=& 
G \biggl [ k_{1}\frac{\omega^{2}}{\Gamma_{0}^{2}+\omega^{2}}
+k_{3}+k_{2} \omega^{2}
\int_{0}^{\infty} \frac{P_{0}(z) dz}{\Gamma_{0}^{2}\exp(-2z)+\omega^{2}} \biggr ],
\nonumber\\
G^{\prime\prime}(\omega) &=& 
G \omega \biggl [ k_{1}\frac{\Gamma_{0}}{\Gamma_{0}^{2}+\omega^{2}}
+k_{2} \int_{0}^{\infty} 
\frac{\Gamma_{0} \exp(-z)P_{0}(z) dz }{\Gamma_{0}^{2}\exp(-2z)+\omega^{2}}\biggr ],
\end{eqnarray}
where $k_{1}=\kappa_{0}K_{1}$, $k_{2}=\kappa_{0}K_{2}$, and
$k_{3}=1-\kappa_{0}$.
The coefficient $k_{1}$ is the concentration of CRRs whose rates of
rearrangement substantially exceeds the maximal frequency of
oscillations (``viscous" units),
$k_{2}$ is the concentration of CRRs whose rates of
rearrangement belong to the frequency window of
dynamic tests (``viscoelastic" units),
and $k_{3}$ is the fraction of CRRs whose rates of rearrangement
is noticeably less than the minimal frequency of vibrations
(``elastic" units).

Equations (11) and (12) are determined by 5 material constants:
\begin{enumerate}
\item
the elastic modulus $G$,

\item
the attempt rate $\Gamma_{0}$,

\item
the average activation energy per CRR $\langle z\rangle$,

\item
the fraction of ``viscous" CRRs $k_{1}$,

\item
the fraction of ``elastic" CRRs $k_{3}$.
\end{enumerate}
These parameters are found by fitting experimental data 
for a multicomponent silicate glass.

\section{Validation of the model}

We study the mechanical response of a rock wool glass
with the chemical composition (wt.-\%) 41.9 SiO$_{2}$, 17.0 Al$_{2}$O$_{3}$,
18.3 CaO, 12.5 MgO, 0.3 K$_{2}$O, 1.4 TiO$_{2}$, 5.7 FeO, 2.0 Na$_{2}$O.
Torsional dynamic tests were performed by N.S. Bagdasarov
(University of Frankfurt, Germany)  at temperatures ranged 
from 500 to 760 $^{\circ}$C (this interval includes the glass transition
point $T_{\rm g}=678$ $^{\circ}$C).
For a description of specimens and the experimental procedure, see \cite{BD98}.
The effect of temperature on the viscielastic behavior of the glass
is demonstrated by Figure 1, where the loss tangent (at the frequency 
$f=1$Hz) is plotted versus temperature.
To approximate observations, we present Eqs. (12) in the form
\begin{eqnarray}
G^{\prime}(\omega) &=& \omega^{2} \biggl [ 
\frac{C_{1}}{\Gamma_{0}^{2}+\omega^{2}}
+\frac{C_{3}}{\omega^{2}}
+C_{2} \int_{0}^{\infty} \frac{P_{0}(z) dz}{\Gamma_{0}^{2}\exp(-2z)+\omega^{2}} \biggr ],
\nonumber\\
G^{\prime\prime}(\omega) &=& 
\omega \biggl [ \frac{C_{1}\Gamma_{0}}{\Gamma_{0}^{2}+\omega^{2}}
+C_{2} \int_{0}^{\infty} 
\frac{\Gamma_{0} \exp(-z)P_{0}(z) dz }{\Gamma_{0}^{2}\exp(-2z)+\omega^{2}}\biggr ],
\end{eqnarray}
where $C_{m}=Gk_{m}$ ($m=1,2,3$).

To find $\Gamma_{0}$, $Z$ and $C_{m}$
at a given temperature, $T$, we fix some intervals 
$[0,\Gamma_{\max}]$ and $[0,Z_{\max}]$, 
where the ``best-fit" parameters $\Gamma_{0}$ and $Z$ 
are assumed to be located,
and divide these intervals into $J$ subintervals by
the points $\Gamma^{(j)}=j\Delta \Gamma$ 
and $Z^{(k)}=k\Delta Z$  ($j,k=1,\ldots,J$)
with $\Delta \Gamma=\Gamma_{\max}/J$ 
and $\Delta Z=Z_{\max}/J$.
For any pair, $\{ \Gamma^{(j)}, Z^{(k)} \}$,
the constants $C_{m}=C_{m}(j,k)$ 
are found by the least-squares technique from the condition 
of minimum of the function
\[ M(j,k) = \sum_{\omega_{l}} \Bigl [ \Bigl (G^{\prime}_{\rm exp}(\omega_{l})
-G^{\prime}_{\rm num}(\omega_{l}) \Bigr )^{2}
+\Bigl (G^{\prime\prime}_{\rm exp}(\omega_{l})
-G^{\prime\prime}_{\rm num}(\omega_{l}) \Bigr )^{2} \Bigr ],
\]
where $G^{\prime}_{\rm exp}$ and $G^{\prime\prime}_{\rm exp}$ are the storage
and loss moduli measured in a test, 
the quantities $G^{\prime}_{\rm num}$ and $G^{\prime\prime}_{\rm num}$ 
are given by Eqs. (13),
and the sum is calculated over all frequencies $\omega_{l}$.
The ``best-fit" parameters $\Gamma_{0}$ and $Z$ minimize the function
$M$ on the set $ \{ \Gamma^{(j)}, Z^{(k)} \quad (j,k=1,\ldots, J)  \}$.
After the constants $C_{m}$ are determined, the elastic modulus, $G$,
and the constants $k_{m}$ are calculated by the formulas
$G=C_{1}+C_{2}+C_{3}$, $k_{m}=C_{m}/G$.
The elastic modulus is plotted versus temperature $T$ in Figure 14,
and the constants, $k_{m}$, are depicted in Figures 15 and 16.
The experimental data are approximated by the linear equations
\begin{equation}
G=G_{0}-G_{1}T, \quad
k_{m}=k_{m0}+k_{m1}T,
\end{equation}
where $G_{j}$ and $k_{mj}$ ($j=0,1$) are found by
the least-squares technique.
The quantities $\langle z\rangle=4Z$ and $\Gamma_{0}$ 
are plotted versus temperature in Figure 17 
together with their mean values $\langle z\rangle=1.35$
and $\Gamma_{0}=0.14$ s$^{-1}$.

\section{Discussion}

Figure 14 shows that $G$ is independent of $T$
in the sub-$T_{\rm g}$ region, and it decreases with $T$ above $T_{\rm g}$.
This conclusion is in accord with the conventional assertion 
that the elastic modulus falls down when the temperature 
reaches the glass transition point.
A decrease in $G$ with temperature is quite noticeable
(the modulus is reduced by 32 \%),
but it is substantially less than that in polymeric glasses 
(where elastic moduli drop by several orders of magnitude
as temperature passes through the glass transition point).

Figures 15 and 16 reveal that the concentration of ``viscous" CRR increases
with temperature both below and above the glass transition point, whereas
the fractions of ``elastic" and viscoelastic relaxing units remain practically
constant below $T_{\rm g}$.
The latter means that only a portion of passive CRRs 
with negligible potential energies become activated in the sub--$T_{\rm g}$
region.
Above the glass transition point, $k_{1}$ and $k_{2}$ substantially
increase with temperature, while $k_{3}$ pronounsly decreases.
This reflects transformation of  passive relaxing units
into active ``viscous" and ``viscoelastic" CRRs induced by thermal fluctuations.

Figure 17 demonstrates that the attept rate, $\Gamma_{0}$, and
the average activation energy of CRRs, $\langle z\rangle$,
are independent of temperature.
The fact that the energy landscape of a glass is not affected 
by temperature appears to be rather natural \cite{WAR00}.
To explain why $\Gamma_{0}$ remains temperature-independent,
we refer to the Narayanaswamy theory \cite{Nar71}.
According to it, rearrangement is CRRs is governed by some
effective temperature, $T_{\rm eff}$, 
\begin{equation}
\frac{1}{T_{\rm eff}}=\frac{\alpha}{T}+\frac{1-\alpha}{T_{\rm f}},
\end{equation}
where $T_{\rm f}$ is the fictive temperature (that characterizes 
changes in the internal structure of a glass), 
and $\alpha\in [0,1]$ is a material constant.
With an increase in temperature, the first term on the right-hand side
of Eq. (15) decreases.
This decrease is, however, compensated by an increase in the other term:
given a duration of a dynamic test, structural recovery in a glass 
occurs more intensively at a higher temperature, which implies 
that a drop in the fictive temperature (that approaches the current 
temperature for an equilibrated glass) is more pronounced 
compared to that at a low temperature.

It is worth to compare Eq. (8) with other micro-mechanical
concepts in viscoelasticity of glasses.
The ADWP (asymmetric double-well potential) model \cite{Buc01} does 
not account for the presence of ``viscous" CRRs that play the key role in
the time-dependent response (Figure 15).
The SGR (soft glassy rheology) model \cite{Sol98} presumes strain-induced 
transformations of the energy landscape that 
are not observed in experiments (Figure 17).

\section{Concluding remarks}

Constitutive equations have been developed for the viscoelastic behavior
of a silicate glass at isothermal uniaxial loading with small strains.
A glass is treated as an ensemble of independent units, where relaxation
of stresses is driven by cooperative rearrangement of clusters of atoms.
The rearrangement process is thought of as a series of hops of CRRs
(trapped at the bottom levels of their potential wells on the energy landscape)
to higher energy levels.
The ensemble is split into three groups of relaxing units: ``elastic" CRRs, 
whose rate of rearrangement is substantially less that
the minimal frequency,
``viscous" CRRs whose rate of rearrangement noticeably exceeds the
maximal frequency, and ``viscoelastic" CRRs, whose
rate of rearrangement is comparable with frequencies of dynamic tests.
Explicit expressions have been derived for the storage and loss 
moduli.
These formulas are determined by 5 adjustable parameters that are found
by matching experimental data for a multicomponent silicate glass at temperatures
in the vicinity of the glass transition point.
It is demonstrated that the attempt rate for hops and the distribution of
potential wells with various depths are independent of temperature,
whereas the instantaneous elastic modulus and the concentrations
of ``elastic", ``viscous" and ``viscoelastic" relaxing units are strongly
affected by temperature.
With an increase in $T$, the fraction of ``elastic" units decreases,
and the concentrations of ``viscous" and ``viscoelastic" units
increase, which is ascribed to thermally-induced activation of passive CRRs.

\setlength{\unitlength}{0.5 mm}
\begin{figure}[t]
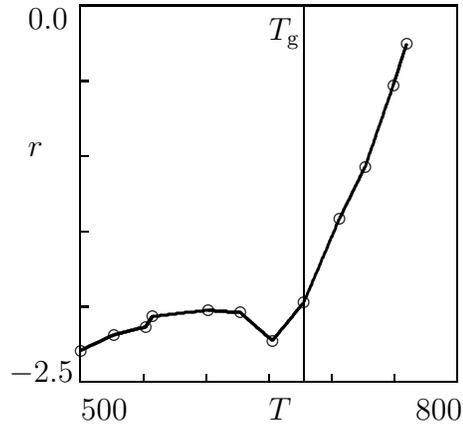

\begin{center}

\end{center}
\vspace*{1 mm}

\caption{The loss tangent $r=\log \tan\delta$ 
versus temperature $T$ $^{\circ}$C.
Circles: treatment of observations.
Solid line: approximation of the experimental data 
by a piecewise linear function.
The vertical line indicates the glass transition temperature}
\end{figure}

\begin{figure}[tbh]
\begin{center}

\end{center}
\vspace*{1 mm}

\caption{The storage modulus $G^{\prime}$ GPa (unfilled circles)
and the loss modulus $G^{\prime\prime}$ GPa (filled circles)
versus frequency $\omega$ rad/s.
Symbols: experimental data ($T=500.0$ $^{\circ}$C).
Solid lines: numerical simulation}
\end{figure}

\begin{figure}[tbh]
\begin{center}
%
\end{center}
\vspace*{1 mm}

\caption{The storage modulus $G^{\prime}$ GPa (unfilled circles)
and the loss modulus $G^{\prime\prime}$ GPa (filled circles)
versus frequency $\omega$ rad/s.
Symbols: experimental data ($T=526.5$ $^{\circ}$C).
Solid lines: numerical simulation}
\end{figure}

\begin{figure}[tbh]
\begin{center}
%
\end{center}
\vspace*{1 mm}

\caption{The storage modulus $G^{\prime}$ GPa (unfilled circles)
and the loss modulus $G^{\prime\prime}$ GPa (filled circles)
versus frequency $\omega$ rad/s.
Symbols: experimental data ($T=552.0$ $^{\circ}$C).
Solid lines: numerical simulation}
\end{figure}

\begin{figure}[tbh]
\begin{center}
%
\end{center}
\vspace*{1 mm}

\caption{The storage modulus $G^{\prime}$ GPa (unfilled circles)
and the loss modulus $G^{\prime\prime}$ GPa (filled circles)
versus frequency $\omega$ rad/s.
Symbols: experimental data ($T=577.0$ $^{\circ}$C).
Solid lines: numerical simulation}
\end{figure}

\begin{figure}[tbh]
\begin{center}
%
\end{center}
\vspace*{1 mm}

\caption{The storage modulus $G^{\prime}$ GPa (unfilled circles)
and the loss modulus $G^{\prime\prime}$ GPa (filled circles)
versus frequency $\omega$ rad/s.
Symbols: experimental data ($T=760.0$ $^{\circ}$C).
Solid lines: numerical simulation}
\end{figure}

\begin{figure}[t]
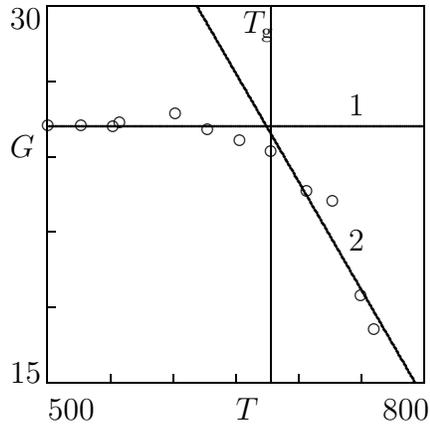

\begin{center}
%
\end{center}
\vspace*{1 mm}

\caption{The elastic modulus $G$ GPa versus temperature 
$T$ $^{\circ}$C.
Circles: treatment of observations.
Solid lines: approximation of the experimental data by Eq. (14).
Curve 1: $G_{0}=25.23$, $G_{1}=0.0$;
curve 2: $G_{0}=83.48$, $G_{1}=8.63\cdot 10^{-2}$}
\end{figure}

\begin{figure}[t]
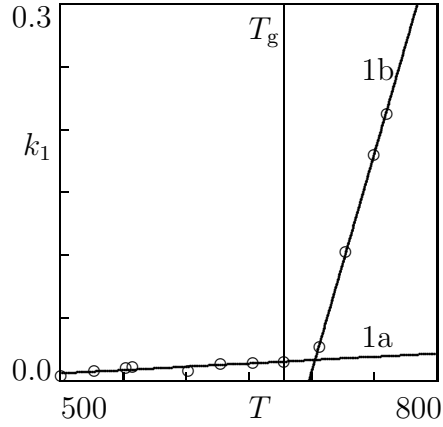

\begin{center}
%
\end{center}
\vspace*{1 mm}

\caption{The dimensionless parameter $k_{1}$ versus temperature 
$T$ $^{\circ}$C.
Circles: treatment of observations.
Solid lines: approximation of the experimental data by Eq. (14).
Curve 1a: $k_{10}=-2.09\cdot 10^{-2}$, $k_{11}=5.31\cdot 10^{-4}$;
curve 1b: $k_{10}=-2.43$, $k_{11}=3.48\cdot 10^{-3}$}
\end{figure}

\begin{figure}[t]
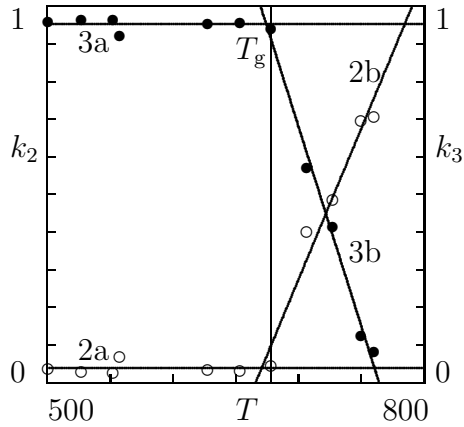

\begin{center}
%
\end{center}
\vspace*{1 mm}

\caption{The dimensionless parameters $k_{2}$ (unfilled circles)
and $k_{3}$ (filled circles) versus temperature $T$ $^{\circ}$C.
Symbols: treatment of observations.
Solid lines: approximation of the experimental data by Eq. (14).
Curve 2a: $k_{20}=3.80\cdot 10^{-2}$, $k_{21}=0.0$;
curve 2b: $k_{20}=-5.34$, $k_{21}=8.02\cdot 10^{-3}$;
curve 3a: $k_{30}=0.95$, $k_{31}=0.0$;
curve 3b: $k_{30}=8.10$, $k_{31}=-1.06\cdot 10^{-2}$}
\end{figure}

\begin{figure}[t]
\begin{center}
\begin{picture}(100,100)
\put(0,0){\framebox(100,100)}
\multiput(16.67,0)(16.67,0){5}{\line(0,1){2}}
\multiput(0,16.67)(0,16.67){5}{\line(1,0){2}}
\multiput(100,12.5)(0,12.5){7}{\line(-1,0){2}}
\put(0,-10){500}
\put(89,-10){800}
\put(50,-10){$T$}
\put(-10,0){0}
\put(-10,94){3}
\put(-13,60){$\langle z\rangle$}
\put(103,0){0}
\put(103,94){4}
\put(103,60){$\Gamma_{0}$}
\put(90,48){1}
\put(90, 6){2}

\put(   8.83,   40.00){\circle{2.5}} 
\put(  17.33,   53.33){\circle{2.5}} 
\put(  42.50,   56.67){\circle{2.5}} 
\put(  51.00,   33.33){\circle{2.5}} 
\put(  59.33,   53.33){\circle{2.5}} 
\put(  75.67,   56.67){\circle{2.5}} 
\put(  83.17,   34.67){\circle{2.5}} 
\put(  86.67,   32.00){\circle{2.5}} 
\multiput(   0.50,   45.00)(0.5,0){200}{\circle*{1.0}} 

\put(   8.83,    2.50){\circle*{2.5}} 
\put(  17.33,    5.00){\circle*{2.5}} 
\put(  19.00,   15.00){\circle*{2.5}} 
\put(  42.50,    7.50){\circle*{2.5}} 
\put(  51.00,    1.25){\circle*{2.5}} 
\put(  59.33,    4.25){\circle*{2.5}} 
\put(  75.67,    3.75){\circle*{2.5}} 
\put(  83.17,    4.00){\circle*{2.5}} 
\put(  86.67,    7.50){\circle*{2.5}} 

\multiput(0.50,3.46)(0.5,0){200}{\circle*{1.0}} 
\end{picture}
\end{center}
\vspace*{1 mm}

\caption{The average activation energy per CRR 
$\langle z\rangle$ (unfilled circles) 
and the attempt rate $\Gamma_{0}$ s$^{-1}$ (filled circles) 
versus temperature $T$ $^{\circ}$C.
Symbols: treatment of observations.
Curves 1 and 2: the mean values}
\end{figure}
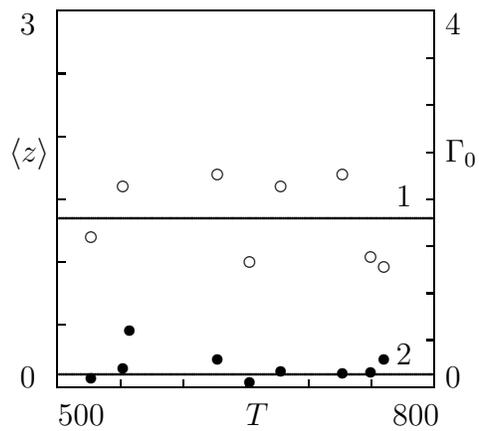

\end{document}